\pdfoutput=1

\documentclass[12pt,a4paper]{article}

\usepackage{ifthen} 
\newboolean{pdflatex}
\setboolean{pdflatex}{true} 

\newboolean{articletitles}
\setboolean{articletitles}{true} 

\newboolean{uprightparticles}
\setboolean{uprightparticles}{false} 

\newboolean{inbibliography}
\setboolean{inbibliography}{false} 

\usepackage{microtype}
\usepackage{lineno}  
\usepackage{xspace} 

\usepackage{caption} 

\usepackage{graphicx}  
\usepackage{color}
\usepackage{colortbl}
\graphicspath{{./}} 

\usepackage{amsmath} 
\usepackage{amssymb}
\usepackage{amsfonts}
\usepackage{upgreek} 

\newcommand*\patchAmsMathEnvironmentForLineno[1]{%
\expandafter\let\csname old#1\expandafter\endcsname\csname #1\endcsname
\expandafter\let\csname oldend#1\expandafter\endcsname\csname
end#1\endcsname
 \renewenvironment{#1}%
   {\linenomath\csname old#1\endcsname}%
   {\csname oldend#1\endcsname\endlinenomath}%
}
\newcommand*\patchBothAmsMathEnvironmentsForLineno[1]{%
  \patchAmsMathEnvironmentForLineno{#1}%
  \patchAmsMathEnvironmentForLineno{#1*}%
}
\AtBeginDocument{%
\patchBothAmsMathEnvironmentsForLineno{equation}%
\patchBothAmsMathEnvironmentsForLineno{align}%
\patchBothAmsMathEnvironmentsForLineno{flalign}%
\patchBothAmsMathEnvironmentsForLineno{alignat}%
\patchBothAmsMathEnvironmentsForLineno{gather}%
\patchBothAmsMathEnvironmentsForLineno{multline}%
}

\usepackage{hyperref}    
\usepackage[all]{hypcap} 

\def\lhcb {LHCb\xspace}
\def\ux85 {UX85\xspace}

\ifthenelse{\boolean{uprightparticles}}%
{

 \def\Pmu         {\ensuremath{\upmu}\xspace}

 \def\Ppi         {\ensuremath{\uppi}\xspace}

 \def\Ppsi        {\ensuremath{\uppsi}\xspace}

 \def\PDelta      {\ensuremath{\Delta}\xspace}                 
 \def\PXi      {\ensuremath{\Xi}\xspace}                 
 \def\PLambda      {\ensuremath{\Lambda}\xspace}                 
 \def\PSigma      {\ensuremath{\Sigma}\xspace}                 
 \def\POmega      {\ensuremath{\Omega}\xspace}                 
 \def\PUpsilon      {\ensuremath{\Upsilon}\xspace}                 
 
 \def\PB      {\ensuremath{\mathrm{B}}\xspace}                 
                  
 \def\PD      {\ensuremath{\mathrm{D}}\xspace}

 \def\PJ      {\ensuremath{\mathrm{J}}\xspace}                 
 \def\PK      {\ensuremath{\mathrm{K}}\xspace}

 \def\Pb      {\ensuremath{\mathrm{b}}\xspace}                 
 \def\Pc      {\ensuremath{\mathrm{c}}\xspace}

 \def\Pi      {\ensuremath{\mathrm{i}}\xspace}

 \def\Ps      {\ensuremath{\mathrm{s}}\xspace}

}
{

 \def\Pmu         {\ensuremath{\mu}\xspace}

 \def\Ppi         {\ensuremath{\pi}\xspace}

 \def\Ppsi        {\ensuremath{\psi}\xspace}                 
                  
 \mathchardef\PDelta="7101
 \mathchardef\PXi="7104
 \mathchardef\PLambda="7103
 \mathchardef\PSigma="7106
 \mathchardef\POmega="710A
 \mathchardef\PUpsilon="7107
                  
 \def\PB      {\ensuremath{B}\xspace}                 
                  
 \def\PD      {\ensuremath{D}\xspace}

 \def\PJ      {\ensuremath{J}\xspace}                 
 \def\PK      {\ensuremath{K}\xspace}

 \def\Pb      {\ensuremath{b}\xspace}                 
 \def\Pc      {\ensuremath{c}\xspace}

 \def\Pi      {\ensuremath{i}\xspace}

 \def\Ps      {\ensuremath{s}\xspace}

}

\def\mun        {\ensuremath{\Pmu^-}\xspace} 

\def\squark    {\ensuremath{\Ps}\xspace}

\def\cquark    {\ensuremath{\Pc}\xspace}

\def\bquark    {\ensuremath{\Pb}\xspace}

\def\pion  {\ensuremath{\Ppi}\xspace}

\def\pip   {\ensuremath{\pion^+}\xspace}

\def\kaon  {\ensuremath{\PK}\xspace}
  \def\Kbar  {\kern 0.2em\overline{\kern -0.2em \PK}{}\xspace}

\def\Kz    {\ensuremath{\kaon^0}\xspace}
\def\Kzb   {\ensuremath{\Kbar^0}\xspace}
\def\KzKzb {\ensuremath{\Kz \kern -0.16em \Kzb}\xspace}
\def\Kp    {\ensuremath{\kaon^+}\xspace}
\def\Km    {\ensuremath{\kaon^-}\xspace}

\def\KpKm  {\ensuremath{\Kp \kern -0.16em \Km}\xspace}

  \def\Dbar    {\kern 0.2em\overline{\kern -0.2em \PD}{}\xspace}
\def\D       {\ensuremath{\PD}\xspace}

\def\Dz      {\ensuremath{\D^0}\xspace}
\def\Dzb     {\ensuremath{\Dbar^0}\xspace}
\def\DzDzb   {\ensuremath{\Dz {\kern -0.16em \Dzb}}\xspace}
\def\Dp      {\ensuremath{\D^+}\xspace}
\def\Dm      {\ensuremath{\D^-}\xspace}

\def\DpDm    {\ensuremath{\Dp {\kern -0.16em \Dm}}\xspace}

\def\B       {\ensuremath{\PB}\xspace}
  \def\Bbar    {\kern 0.18em\overline{\kern -0.18em \PB}{}\xspace}

\def\Bub     {\ensuremath{\B^-}\xspace}

\def\Bm      {\ensuremath{\Bub}\xspace}

\def\Bsb     {\ensuremath{\Bbar^0_\squark}\xspace}

\def\jpsi     {\ensuremath{{\PJ\mskip -3mu/\mskip -2mu\Ppsi\mskip 2mu}}\xspace}

  \def\Y#1S{\ensuremath{\PUpsilon{(#1S)}}\xspace}

\def\to                 {\ensuremath{\rightarrow}\xspace}

\def\AT#1     {\ensuremath{A_{\mathrm{T}}^{#1}}\xspace}           

\def\C#1      {\ensuremath{\mathcal{C}_{#1}}\xspace}                       
\def\Cp#1     {\ensuremath{\mathcal{C}_{#1}^{'}}\xspace}                    
\def\Ceff#1   {\ensuremath{\mathcal{C}_{#1}^{\mathrm{(eff)}}}\xspace}        
\def\Cpeff#1  {\ensuremath{\mathcal{C}_{#1}^{'\mathrm{(eff)}}}\xspace}       
\def\Ope#1    {\ensuremath{\mathcal{O}_{#1}}\xspace}                       
\def\Opep#1   {\ensuremath{\mathcal{O}_{#1}^{'}}\xspace}                    

\newcommand{\tev}{\ensuremath{\mathrm{\,Te\kern -0.1em V}}\xspace}
\newcommand{\gev}{\ensuremath{\mathrm{\,Ge\kern -0.1em V}}\xspace}
\newcommand{\mev}{\ensuremath{\mathrm{\,Me\kern -0.1em V}}\xspace}
\newcommand{\kev}{\ensuremath{\mathrm{\,ke\kern -0.1em V}}\xspace}
\newcommand{\ev}{\ensuremath{\mathrm{\,e\kern -0.1em V}}\xspace}
\newcommand{\gevc}{\ensuremath{{\mathrm{\,Ge\kern -0.1em V\!/}c}}\xspace}
\newcommand{\mevc}{\ensuremath{{\mathrm{\,Me\kern -0.1em V\!/}c}}\xspace}
\newcommand{\gevcc}{\ensuremath{{\mathrm{\,Ge\kern -0.1em V\!/}c^2}}\xspace}
\newcommand{\gevgevcccc}{\ensuremath{{\mathrm{\,Ge\kern -0.1em V^2\!/}c^4}}\xspace}
\newcommand{\mevcc}{\ensuremath{{\mathrm{\,Me\kern -0.1em V\!/}c^2}}\xspace}

\def\mum  {\ensuremath{\,\upmu\rm m}\xspace}

\def\invfb   {\ensuremath{\mbox{\,fb}^{-1}}\xspace}

\def\sec  {\ensuremath{\rm {\,s}}\xspace}

\def\gsim{{~\raise.15em\hbox{$>$}\kern-.85em
          \lower.35em\hbox{$\sim$}~}\xspace}
\def\lsim{{~\raise.15em\hbox{$<$}\kern-.85em
          \lower.35em\hbox{$\sim$}~}\xspace}

\def\pt         {\mbox{$p_{\rm T}$}\xspace}

\def\evtgen     {\mbox{\textsc{EvtGen}}\xspace}
\def\pythia     {\mbox{\textsc{Pythia}}\xspace}

\def\geant      {\mbox{\textsc{Geant4}}\xspace}

\def\gauss      {\mbox{\textsc{Gauss}}\xspace}

\def\tell1  {TELL1\xspace}
\def\ukl1   {UKL1\xspace}

\newcommand{\etal}{{\slshape et al.\/}\xspace}

\usepackage{cite} 
\usepackage{setspace}
\usepackage{mciteplus}

\begin{document}

\renewcommand{\thefootnote}{\fnsymbol{footnote}}
\setcounter{footnote}{1}

\begin{titlepage}
\pagenumbering{roman}

\vspace*{-1.5cm}
\centerline{\large EUROPEAN ORGANIZATION FOR NUCLEAR RESEARCH (CERN)}
\vspace*{1.5cm}
\hspace*{-5mm}\begin{tabular*}{16cm}{lc@{\extracolsep{\fill}}r}
\vspace*{-12mm}\mbox{\!\!\!\includegraphics[width=.12\textwidth]{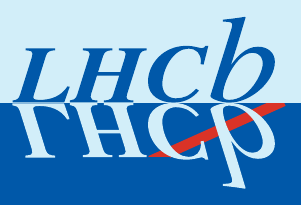}}& & \\ 
& & CERN-PH-EP-2014-002\\
& & LHCb-PAPER-2013-064\\  
& &\today \\ 
\end{tabular*}

\vspace*{3.0cm}

{\bf\boldmath\huge
\begin{center}
Search for Majorana neutrinos in $\Bm \to \pip\mun\mun$ decays 
\end{center}
}

\vspace*{2.0cm}

\begin{center}
The LHCb collaboration\footnote{Authors are listed on the following pages.}
\end{center}

\vspace{\fill}

\begin{abstract}
  \noindent
A search for heavy Majorana neutrinos produced in the $\Bm \to \pip\mun\mun$ decay mode is performed using 3 \invfb  of integrated luminosity collected with the LHCb detector in $pp$ collisions at center-of-mass energies of 7 TeV and 8 TeV at the LHC. Neutrinos with masses in the range $250-5000$~MeV and  lifetimes from zero to 1000~ps are probed. In the absence of a signal, upper limits are set on
 the branching fraction ${\cal{B}}(\Bm \to \pip\mun\mun)$ as functions of neutrino mass and lifetime. These limits are on the order of $10^{-9}$  for short neutrino lifetimes of 1~ps or less. Limits are also set on the coupling between the muon and a possible fourth-generation neutrino. 
\end{abstract}

\vspace*{2.0cm}

\begin{center}
  Submitted to Phys.~Rev.~Lett. 
\end{center}

\vspace{\fill}

{\footnotesize 
\centerline{\copyright~CERN on behalf of the \lhcb collaboration, license \href{http://creativecommons.org/licenses/by/3.0/}{CC-BY-3.0}.}}
\vspace*{2mm}

\end{titlepage}

\newpage
\setcounter{page}{2}
\mbox{~}

\centerline{\large\bf LHCb collaboration}
\begin{flushleft}
\small
R.~Aaij$^{40}$, 
B.~Adeva$^{36}$, 
M.~Adinolfi$^{45}$, 
A.~Affolder$^{51}$, 
Z.~Ajaltouni$^{5}$, 
J.~Albrecht$^{9}$, 
F.~Alessio$^{37}$, 
M.~Alexander$^{50}$, 
S.~Ali$^{40}$, 
G.~Alkhazov$^{29}$
% ...

{\footnotesize \it
% ...
$ ^{5}$Clermont Universit\'{e}, Universit\'{e} Blaise Pascal, CNRS/IN2P3, LPC, Clermont-Ferrand, France\\
% ...
$ ^{9}$Fakult\"{a}t Physik, Technische Universit\"{a}t Dortmund, Dortmund, Germany\\
% ...
$ ^{29}$Petersburg Nuclear Physics Institute (PNPI), Gatchina, Russia\\
% ...
$ ^{36}$Universidad de Santiago de Compostela, Santiago de Compostela, Spain\\
$ ^{37}$European Organization for Nuclear Research (CERN), Geneva, Switzerland\\
% ...
$ ^{40}$Nikhef National Institute for Subatomic Physics, Amsterdam, The Netherlands\\
% ...
$ ^{45}$H.H. Wills Physics Laboratory, University of Bristol, Bristol, United Kingdom\\
% ...
$ ^{50}$School of Physics and Astronomy, University of Glasgow, Glasgow, United Kingdom\\
$ ^{51}$Oliver Lodge Laboratory, University of Liverpool, Liverpool, United Kingdom\\
% ...
}
\end{flushleft}

\cleardoublepage

\renewcommand{\thefootnote}{\arabic{footnote}}
\setcounter{footnote}{0}



\pagestyle{plain} 
\setcounter{page}{1}
\pagenumbering{arabic}  
\noindent 

\begin{figure}[b!]
\begin{center}
\includegraphics[width=3.9 in]{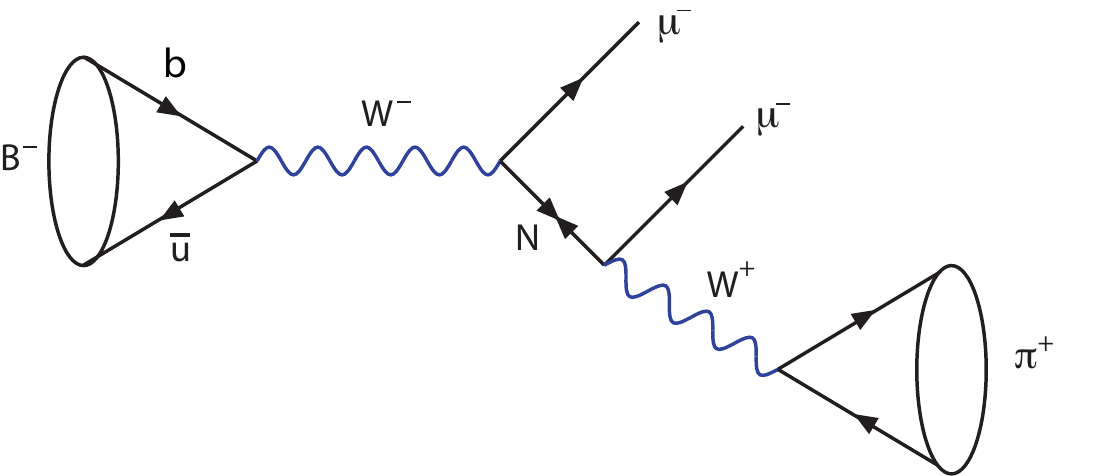}
\end{center}
\vspace{-10pt}
\caption{Feynman diagram for $\Bm\to\pi^+\mu^-\mu^-$ decay via a Majorana neutrino labelled $N$.} 
\label{fig:Majorana-feyn}
\end{figure}

Neutrinos can either be their own anti-particles, in which case they are called ``Majorana" particles \cite{Majorana}, or Dirac fermions.  Heavy Majorana neutrinos can be sought in heavy flavor decays and couplings to a single fourth neutrino generation can be determined, or limits imposed, as in previous measurements \cite{Aaij:2012zr,*Aaij:2011ex,Lees:2013pxa,*Seon:2011ni,*Weir:1989sq}. The lepton number violating process $B^-\to\pi^+\mu^-\mu^-$, shown in Fig.~\ref{fig:Majorana-feyn}, is forbidden in the Standard Model, but  can proceed via the production of on-shell Majorana neutrinos. It is one of the most sensitive ways of looking for these particles in $B$ meson decays, and has been modeled by Atre \etal \cite{Atre:2009rg}.  Note that it is possible for virtual Majorana neutrinos of any mass to contribute to this decay.   We investigate the $B^-\to\pi^+\mu^-\mu^-$ decay using 3~fb$^{-1}$ of data acquired by the LHCb experiment in $pp$ collisions. (In this Letter mention of a particular decay implies the use of the charge-conjugate decay as well.) One third of the data was recorded at 7 TeV center-of-mass energy, and the remainder at 8 TeV.  

The search strategy is based on our previous analysis \cite{Aaij:2012zr},  but extends the sensitivity 
to neutrino lifetimes, $\tau_N$, from the picosecond range up to about 1000~ps. The selection is aimed at maximizing the efficiency squared divided by the background yield.  For lifetimes $\gtrsim$1~ps, the $\pi^+\mu^-$ decay products can appear as significantly detached from the $B^-$ decay vertex. Therefore, we use two distinct strategies, one for short $\tau_N$ ($\cal{S}$) and another for $\tau_N$  up to 1000 ps ($\cal{L}$).
 
The \lhcb detector is a single-arm forward
spectrometer covering the \mbox{pseudorapidity} range $2<\eta <5$, designed
for the study of particles containing \bquark or \cquark quarks \cite{LHCb-det}. The detector includes a high precision tracking system consisting of a silicon-strip vertex detector surrounding the $pp$ interaction region (VELO),
a large-area silicon-strip detector located upstream of a dipole
magnet with a bending power of about $4{\rm\,Tm}$, and three stations
of silicon-strip detectors and straw drift tubes placed
downstream. The combined tracking system provides a momentum ($p$) measurement with
relative uncertainty that varies from 0.4\% at 5~GeV to 0.6\% at 100~GeV. (We use natural units where $c$=1.)
The impact parameter (IP) is defined as the minimum track distance with respect to the primary vertex (PV). For tracks with large transverse momentum (\pt) with respect to the proton beam direction, the IP resolution is approximately 20\mum. Charged hadrons are identified using two ring-imaging Cherenkov (RICH) detectors. Photon, electron and hadron candidates are identified by a calorimeter system consisting of scintillating-pad and pre-shower detectors, an electromagnetic calorimeter and a hadronic calorimeter. Muons are identified by a system composed of alternating layers of iron and multiwire proportional chambers. 

 The LHCb trigger \cite{Aaij:2012me} consists of a hardware stage, based on information from the
calorimeter and muon systems, followed by a software stage that performs full event reconstruction. The hardware trigger selects either a  single muon candidate with $\pt>1.64$~GeV, or two muons with the product of their \pt values being greater than 1.69~GeV$^2$.  In the subsequent software stage  a muon candidate must form a vertex with one or two additional tracks that are detached from the PV.  The trigger efficiency decreases for large lifetime Majorana neutrinos. 
Simulations are performed using \pythia \cite{Sjostrand:2006za} with the specific
tuning given in Ref. \cite{LHCb-PROC-2011-005}, and the LHCb detector description based on \geant \cite{Allison:2006ve,*Agostinelli:2002hh}
described in Ref. \cite{LHCb-PROC-2011-006}. Decays of $b$ hadrons are based on \evtgen \cite{Lange:2001uf}. Simulation of $\Bm\to\pi^+\mu^-\mu^-$ is carried out in two steps, the first being the two body decay $\Bm\to N\mu^-$, where $N$ is a putative Majorana neutrino, and the second $N\to\pi^+\mu^-$.

 In both categories $\cal{S}$ and $\cal{L}$,  only tracks that start in the VELO are used. We require muon candidates to have $p>3$~GeV and $\pt>0.75$~GeV, as muon detection provides fewer fakes above these values.  The hadron must have $p>2$~GeV and $\pt>1.1$~GeV, in order to be tracked well.  Muon candidate tracks are required to have hits in the muon chambers. The same criteria apply for the channel we use for normalization purposes, $B^-\to\jpsi K^-$ with $\jpsi\to\mu^+\mu^-$. Pion and kaon candidates must be positively identified in the RICH systems.  
 For the $\cal{S}$ case and the normalization channel,  candidate \Bm combinations must form a common vertex with a $\chi^2$ per number of degrees of freedom (ndf) less than 4. For the $\cal{L}$ candidates we require that the $\pi^+\mu^-$ tracks form a neutrino candidate ($N$) decay vertex with a $\chi^2<10$. A \Bm candidate decay vertex is searched for by extrapolating the $N$ trajectory back to a near approach with another $\mu^-$ candidate, which 
must form a vertex with the other muon having a $\chi^2 <4$. The distance between the $\pi^+\mu^-$ and the primary vertex divided by its uncertainty must be greater than 10. The \pt of the $\pi^+\mu^-$ pair must also exceed 700~MeV. For both $\cal{S}$ and $\cal{L}$ cases, we require that the cosine of the angle between the \Bm candidate momentum vector and the line from the PV to the \Bm vertex  be greater than 0.99999. 
  The two cases are not exclusive, with 16\% of the event candidates appearing in both.

\begin{figure}[t]
\begin{center}
\includegraphics[width=6.0 in]{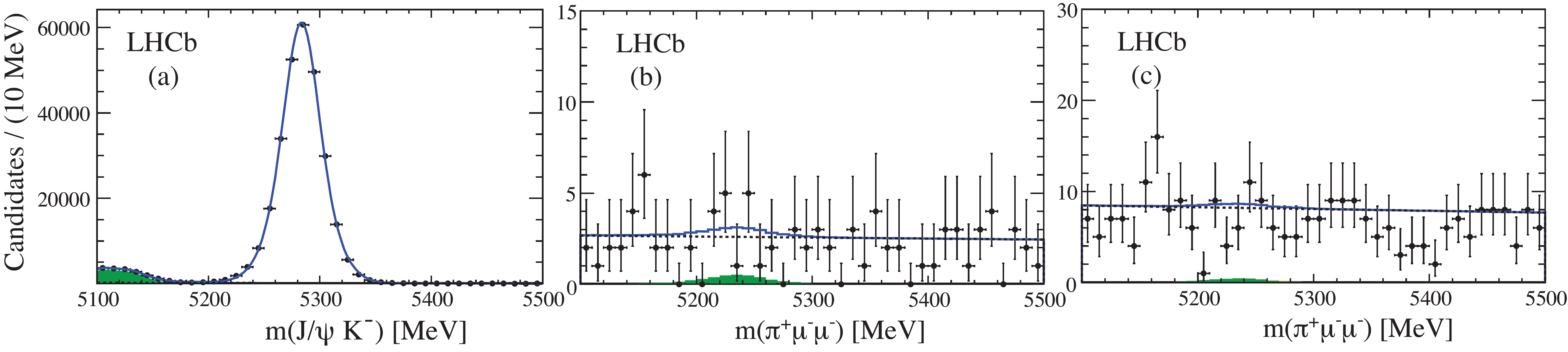}
\end{center}
\vspace{-15pt}
\caption{Invariant mass distributions with fits overlaid of candidate mass spectra for  (a) $\jpsi K^-$, (b) $\pip\mun\mun$ ($\cal{S}$) , and (c) $\pip\mun\mun$ ($\cal{L}$). Backgrounds are (green) shaded; they peak under the signal in (b) and (c).
 The dotted lines show the combinatorial backgrounds only. The solid line shows the sum of both backgrounds.} 
\label{fig:zero_B_mass_fit}
\end{figure}

The mass spectra of the selected candidates are shown in Fig.~\ref{fig:zero_B_mass_fit}. An extended unbinned likelihood fit is performed to the  $\jpsi K^-$ mass spectrum with a double-Crystal Ball function \cite{Skwarnicki:1986xj} plus a triple-Gaussian background to account for partially reconstructed $B$ decays and a linear function for combinatoric background. We find $282\,774\pm543$ signal events in the normalization channel. Backgrounds in the $\pi^+\mu^-\mu^-$ final state come from $B$ decays to charmonium and combinatoric sources. Charmonium  backgrounds are estimated using fully reconstructed $\jpsi K^-(\pi^-)$ and  $\psi(2S) K^-(\pi^-)$ events and are indicated by shaded regions; they can peak at the \Bm mass. No signal is observed in either the ${\cal{S}}$ or ${\cal{L}}$ samples.

We use the  $\text{CL}_{\text{s}}$ method to set upper limits \cite{CLS}, which requires the  determination of the expected background yields and total number of events in the signal region. We define the signal region as the mass interval within $\pm 2\sigma$ of the \Bm mass where $\sigma$ is the mass resolution, specifically $5238.6 - 5319.8$~MeV. Peaking background shapes and normalizations are fixed from exclusive reconstructions in data.
  We fit the distributions outside of the \Bm signal region with a sum of the peaking background tails, where both shape and normalization are fixed, and linear functions to account for the combinatorial backgrounds. The interpolated combinatoric background in each signal region is combined with the peaking background to determine the total background. 

 In the signal $B$ mass range there are 19 events  in the $\cal{S}$ sample and 60 events in the $\cal{L}$  sample.  The $\cal{S}$ and $\cal{L}$ background fit yields are 17.8$\pm$3.2, and 54.5$\pm$5.4, respectively,  in the same region.

The detection efficiency varies as a function of neutrino mass, $m_N$, and changes for the $\cal{L}$ sample with  $\tau_N$. To quote an upper limit on the branching fraction for the $\cal{S}$ sample we take the average detection efficiency, as determined by simulation, with respect to the normalization mode of 0.687$\pm$0.001.  In computing the limit we include 
the uncertainties on background yields obtained from the fit to the $m(\pip\mun\mun)$ distribution, and the systematic uncertainty described below.
The normalization is obtained from the number of $\jpsi K^-$ events and the known rate of ${\cal{B}}(\B^-\to\jpsi K^-,~\jpsi\to\mu^+\mu^-)=(6.04 \pm 0.26) \times 10^{-5}$ \cite{Aaij:2013orb,PDG2012}.  We find
\begin{equation}
\nonumber
 {\cal{B}}(\Bm \to \pip \mun \mun) < 4.0 \times 10^{-9} \rm{~at~}95\%\rm{~confidence~level~( C.L.)}
 \end{equation} 
 This limit is applicable for $\tau_N$ $\lesssim 1$~ps.  The total systematic uncertainty is 6.6\%.
 The largest source  is ${\cal{B}}(B^-\to\jpsi K^-)$ (4.2\%), followed by modeling of the efficiency ratio (3.5\%) and backgrounds (3.5\%), relative particle identification efficiencies (0.5\%),  tracking efficiency differences for kaons versus  pions (0.5\%), and yield of the normalization channel (0.4\%).

\begin{figure}[t]
\begin{center}
\includegraphics[width=6.0 in]{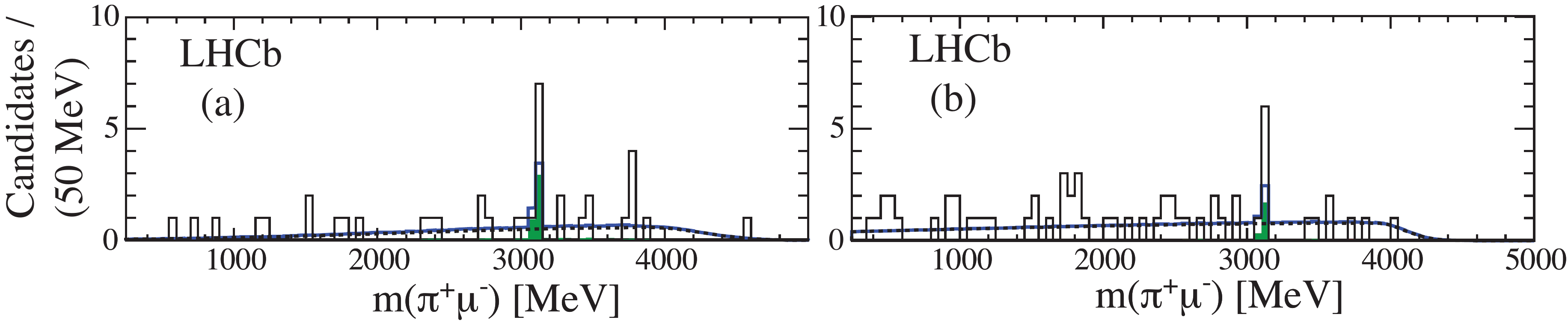}
\end{center}
\vspace{-20pt}
\caption{Invariant $\pip \mun$ mass distribution for  $\pip\mun\mun$ candidates with masses restricted to  $\pm2\sigma$ of $\Bm$ mass for the (a) $\cal{S}$ and (b) $\cal{L}$ selections. The shaded regions indicate the estimated peaking backgrounds. Backgrounds that peak under the signal in (a) and (b) are (green) shaded. The dotted lines show the combinatorial backgrounds only. The solid line the sum of both backgrounds. (In (a) there are two combinations per event.)} 
\label{fig:detached_narrow_B_mass}
\end{figure}

We also search for signals as a function of $m_N$. The $\pi^+\mu^-$ mass spectra are shown in Fig.~\ref{fig:detached_narrow_B_mass} for both $\cal{S}$ and $\cal{L}$ selections, requiring that  the $\pi^+\mu^-\mu^-$ mass be restricted to the  \Bm signal range. There is an obvious peak around 3100~MeV from misidentified $\jpsi K^-$ (or $\pi^-$) events.
The $\pi^+\mu^-$ mass spectra are fitted with a function derived from fitting the upper \Bm sideband regions, from $5319.8-5400.0$~MeV, for the combinatoric background, and  peaking background components obtained from simulation.

\begin{figure}[b]
\begin{center}
\includegraphics[width=3.9 in]{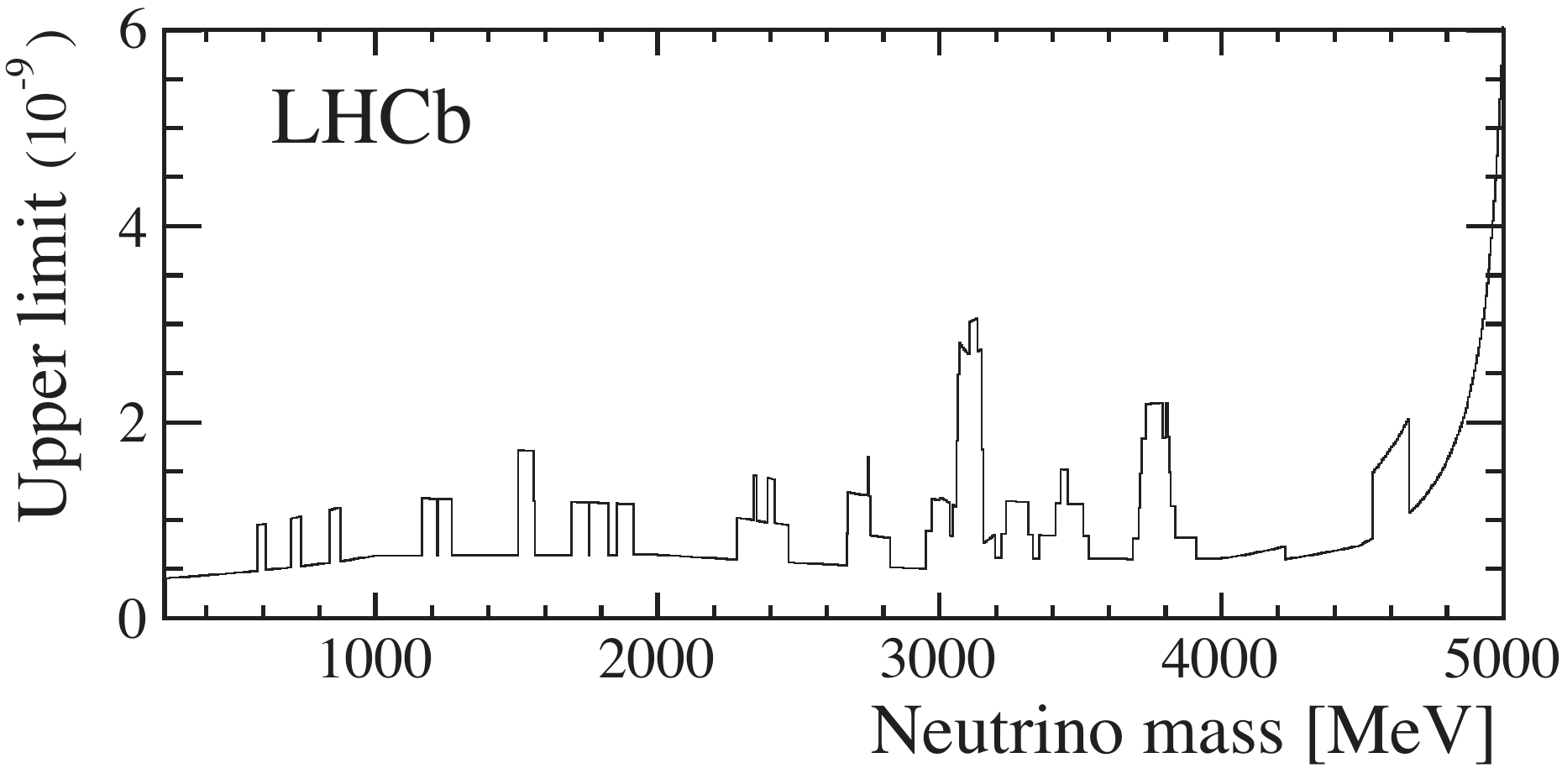}
\end{center}
\vspace{-15pt}
\caption{Upper limit on ${\cal{B}}(\Bm \to \pip \mun \mun)$ at $95\%$ C.L. as a function of $m_N$ in 5~MeV intervals for $\cal{S}$  selected events.} 
\label{fig:upper_lismit_scan}
\end{figure}

As there is no evidence for a signal, upper limits are set by scanning across the $m_N$ spectrum. At every 5 MeV step beginning at 
250~MeV and ending at 5000 MeV we define a $\pm 3\sigma$ search region, where $\sigma$ ranges from approximately 3~MeV at low mass to 24~MeV at high mass. The mass resolution is determined from fitting signals in other LHCb data \cite{Aaij:2012zr}. The fitted background is then subtracted from the event yields in each interval. The upper limit at $95\%$ C.L. of ${\cal{B}}(\Bm \to \pip \mun \mun)$  at each mass value is computed using the CL$_s$ method.  The simulated efficiency ratio to the normalization mode averages about 0.8 up to 4000 MeV, and then approaching the phase space boundary, sharply decreases to 0.2 at 5000 MeV. The results of this scan are shown in Fig.~\ref{fig:upper_lismit_scan}.

\begin{figure}[b]
\begin{center}
\vspace{-5mm}
\includegraphics[width=4.9 in]{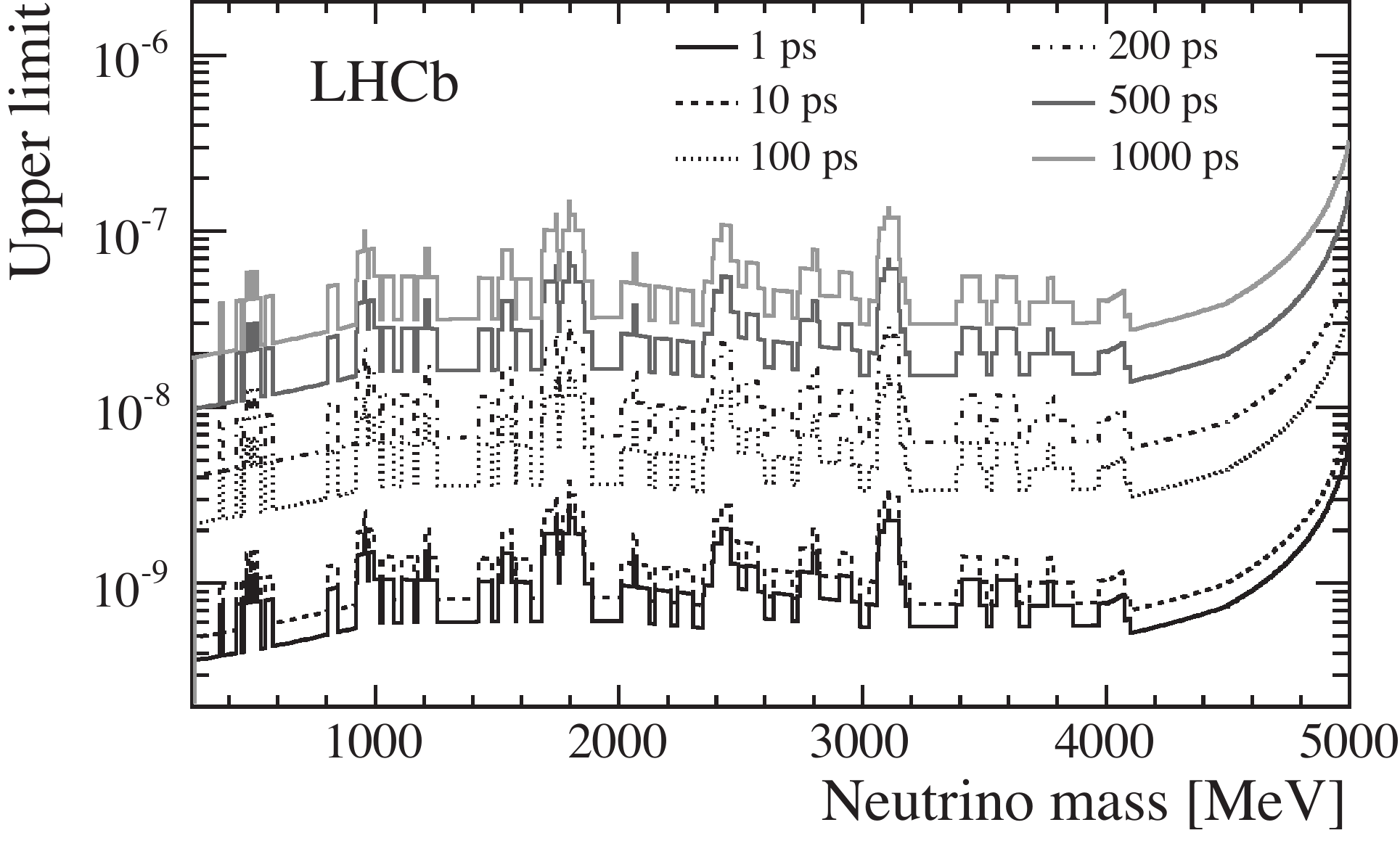}
\end{center}
\vspace{-20pt}
\caption{Upper limits on ${\cal{B}}(\Bm \to \pip \mun \mun)$ at 95\% C.L.  as a function of $m_N$,  in 5~MeV intervals, for specific values of $\tau_N$.} 
\label{fig:UL-mass-lifetime}
\end{figure}

The efficiency is highest for $\tau_N$ of a few ps, and decreases rapidly until about 200 ps when it levels off until about 1000 ps, beyond which it slowly vanishes as most of the decays occur outside of the vertex detector.
For $\cal{L}$  candidates, we set upper limits as a function of both $m_N$ and lifetime by performing the same scan in mass as before, but applying efficiencies appropriate for individual lifetime values between 1 and 1000~ps.  The number of background events is extracted from the sum of combinatorial and peaking backgrounds in the fit to the $m(\pip\mun)$  distribution in the same manner as for the  $\cal{S}$ sample. The estimated signal yield is the difference between the total number of events computed by counting the number in the interval and the fitted background yield. We take the $\tau_N$ dependence into account by using different efficiencies for each lifetime step. 
The two-dimensional plot of the upper limit on ${\cal{B}}(\Bm \to \pip \mun \mun)$, computed using the $\text{CL}_{\text{s}}$ method, is shown in Fig.~\ref{fig:UL-mass-lifetime}. 

Model dependent upper limits on the coupling of a single fourth-generation Majorana neutrino to muons, $|V_{\mu 4}|$, for each value of $m_N$  are extracted using the formula from Atre \etal \cite{Atre:2009rg}
 \begin{equation}
 \label{Eq:BrV}
{\cal{B}}(\Bm \to \pip \mun \mun)=\frac{G_{F}^4 f_{B}^2 f_{\pi}^2m_{B}^{5}}{128 \pi^2\hbar}  
{|V_{ub}V_{ud}|^2 }\tau_B \left(1-\frac{m^2_N}{m_B^2}\right) \frac{m_N}{\Gamma_N}|V_{\mu4}|^{4}, 
\end{equation}
 with $G_{F}=1.166377\times10^{-5} \gev^{-2}$, $f_{B}=0.19 \gev$, $f_{\pi}=0.131 \gev$, $|V_{ub}|=0.004$, $|V_{ud}|=0.9738$, $m_{B}=5.279 \gev$, $\tau_B=1.671$~ps, and $\hbar=6.582\times 10^{-25} \gev\,\sec$ \cite{PDG2012}. The  total neutrino decay width, $\Gamma_N$, is a function of $m_N$ and proportional to $|V_{\mu 4}|^2$. In order to set limits on $|V_{\mu4}|^{2}$ a model for $\Gamma_N$ is required. The purely leptonic modes are specified in Ref.~\cite{Atre:2009rg}. For the hadronic modes we use the fraction of times the charged current manifests itself as a single charged pion in $\tau^-$ and \Bm decays, giving an additional $m^3_N$ dependent factor in $\Gamma_N$. The total width for Majorana neutrino decay then is
 \begin{equation}
 \Gamma_N=\left[3.95 m^3_N+2.00 m^5_N(1.44m^3_N+1.14)\right]10^{-13}|V_{\mu 4}|^2,
 \end{equation}
where $m_N$  and $\Gamma_N$  are in units of GeV.
The first term corresponds to fully leptonic three-body decays, while the second is for decays into one lepton and hadrons. 
 
\begin{figure}[b]
\begin{center}
\includegraphics[width=5.9 in]{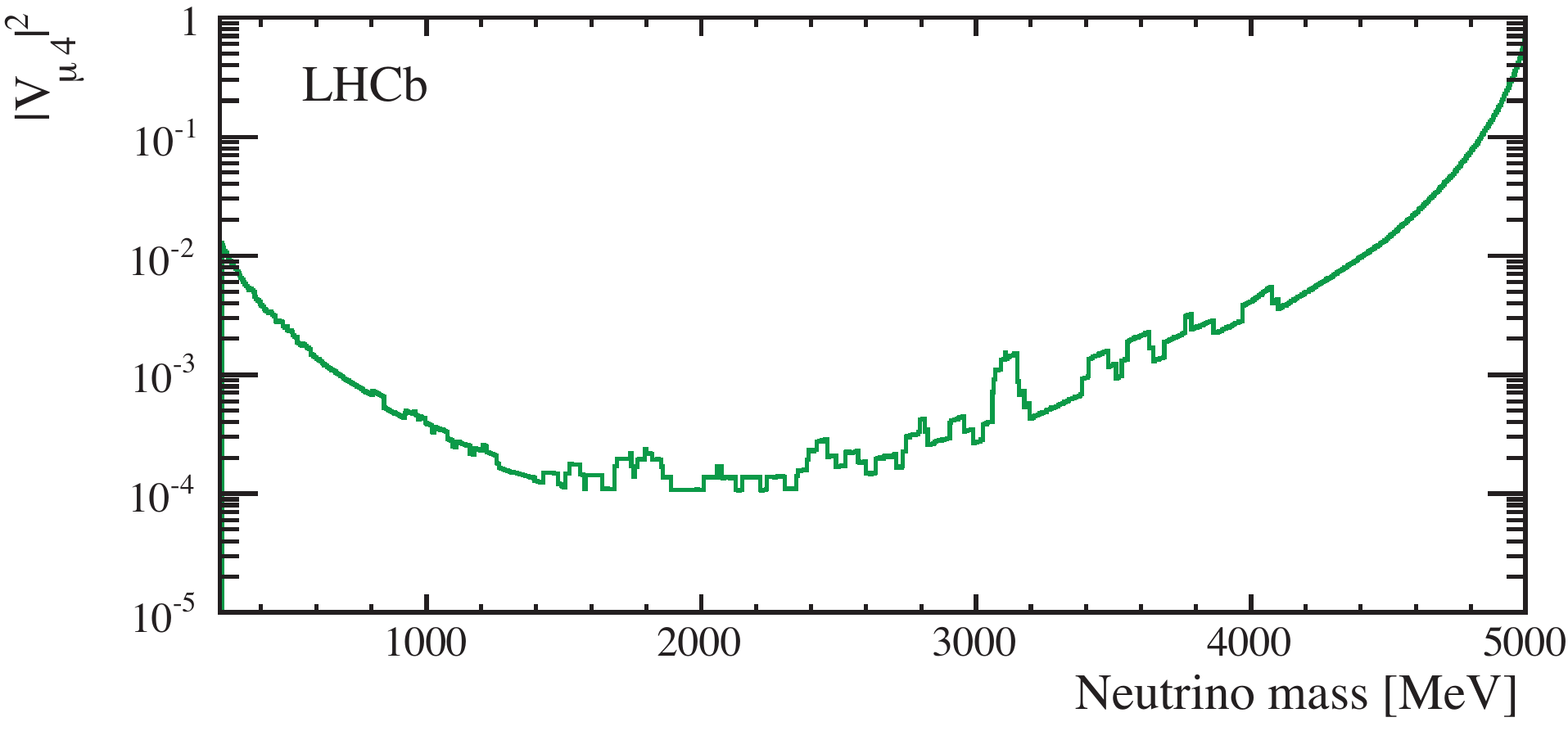}
\end{center}
\vspace{-4mm}
\caption{Upper limits at 95\% C.L. on $|V_{\mu4}|^2$ are shown as a function of  $m_N$ for ${\cal{L}}$ events.} 
\label{fig:Vmu4modified}
\end{figure}

 To obtain upper limits on $|V_{\mu4}|^2$ for each value of $m_N$ we assume a value for  $|V_{\mu4}|$, and calculate $\Gamma_N$. This allows us to determine the $\tau_N$ dependent detection efficiency.  We then use Eq.~(\ref{Eq:BrV}) to find the branching fraction. The value of $|V_{\mu4}|$ is adjusted to match the previously determined upper limit value (see Fig.~\ref{fig:UL-mass-lifetime}).
 The resulting 95\% C.L. limit on $|V_{\mu4}|^2$ is shown in Fig.~\ref{fig:Vmu4modified} as a function of $m_N$. Limits have been derived  by Atre \etal \cite{Atre:2009rg} for other experiments using different assumptions about the dependence of $\Gamma_N$ with $m_N$, and thus cannot be directly compared. 
 More searches exist for higher mass neutrinos \cite{Chatrchyan:2012fla,*Aad:2011vj,*ATLAS:2012ak}.
The results presented here supersede previous LHCb results \cite{Aaij:2012zr,*Aaij:2011ex}, significantly improve the limits on the $\B^-\to\pi^+\mu^-\mu^-$ branching fraction and extend the lifetime range of the Majorana neutrino search from a few picoseconds to one nanosecond. 

In conclusion, we have searched for on-shell Majorana neutrinos coupling to muons in the $\Bm \to \pip \mun \mun$ decay channel as a function of $m_N$ between $250-5000$~MeV and for lifetimes up to $\approx$1000~ps. In the absence of a significant signal, we set upper limits on the $\Bm \to \pip \mun \mun$  branching fraction and the coupling  $|V_{\mu 4}|^2$ as a function of the neutrino mass.  

We express our gratitude to our colleagues in the CERN
accelerator departments for the excellent performance of the LHC. We
thank the technical and administrative staff at the LHCb
institutes. We acknowledge support from CERN and from the national
agencies: CAPES, CNPq, FAPERJ and FINEP (Brazil); NSFC (China);
CNRS/IN2P3 and Region Auvergne (France); BMBF, DFG, HGF and MPG
(Germany); SFI (Ireland); INFN (Italy); FOM and NWO (The Netherlands);
SCSR (Poland); MEN/IFA (Romania); MinES, Rosatom, RFBR and NRC
``Kurchatov Institute'' (Russia); MinECo, XuntaGal and GENCAT (Spain);
SNSF and SER (Switzerland); NAS Ukraine (Ukraine); STFC (United
Kingdom); NSF (USA). We also acknowledge the support received from the
ERC under FP7. The Tier1 computing centres are supported by IN2P3
(France), KIT and BMBF (Germany), INFN (Italy), NWO and SURF (The
Netherlands), PIC (Spain), GridPP (United Kingdom). We are indebted to the communities behind the multiple open source software packages we depend on. We are also thankful for the computing resources and the access to software R\&D tools provided by Yandex LLC (Russia).

\newpage
\addcontentsline{toc}{section}{References}
\ifx\mcitethebibliography\mciteundefinedmacro
\PackageError{LHCb.bst}{mciteplus.sty has not been loaded}
{This bibstyle requires the use of the mciteplus package.}\fi
\providecommand{\href}[2]{#2}

\end{document}